\documentclass[final,5p,times,twocolumn]{elsarticle}

\usepackage{amssymb}
\usepackage{txfonts}
\usepackage{url}
\usepackage{lineno}
\usepackage{amsmath}

\usepackage{hyperref}
\usepackage{xcolor}

\makeatletter
\def\ps@pprintTitle{%
  \let\@oddhead\@empty
  \let\@evenhead\@empty
  \let\@oddfoot\@empty
  \let\@evenfoot\@oddfoot
}
\makeatother

\begin{document}

\begin{frontmatter}

\title{Performance of the Pair Spectrometer in Hall D at Jefferson Lab }

\author[jlab]{A. Somov\corref{cor1}}
\ead{somov@jlab.org}
\author[mephi]{S. Somov}
\author[jlab]{V.V. Berdnikov}
\address[jlab]{Thomas Jefferson National Accelerator Facility, Newport News, VA 23606, USA}
\address[mephi]{National Research Nuclear University MEPhI, Moscow, Russia}

\cortext[cor1]{Corresponding author. Tel.: +1 757 269 5553.}

\begin{abstract}
This article describes the performance of the pair spectrometer installed in experimental Hall D at Jefferson Lab and its operation in multiple experiments with the GlueX detector. The primary purpose of the pair spectrometer is the precise determination of the flux of beam photons incident on the GlueX target, a critical input for physics analyses such as absolute cross-section measurements. The photon energy spectrum is determined by reconstructing electron–positron pairs produced in a thin converter inserted into the photon beam. The spectrometer is integrated into the GlueX trigger system, enabling continuous real-time monitoring of the photon flux and the recording of $e^+e^-$ pair candidates for offline analysis. In addition, the pair spectrometer provides a versatile test facility for evaluating calorimeter prototypes using leptons with well-defined energies produced via electromagnetic pair production.

\end{abstract}


\end{frontmatter}

\section{Introduction}
\label{sec:intro}

\begin{figure*}[t]
\begin{center}
\includegraphics[width=1.0\linewidth,angle=0]{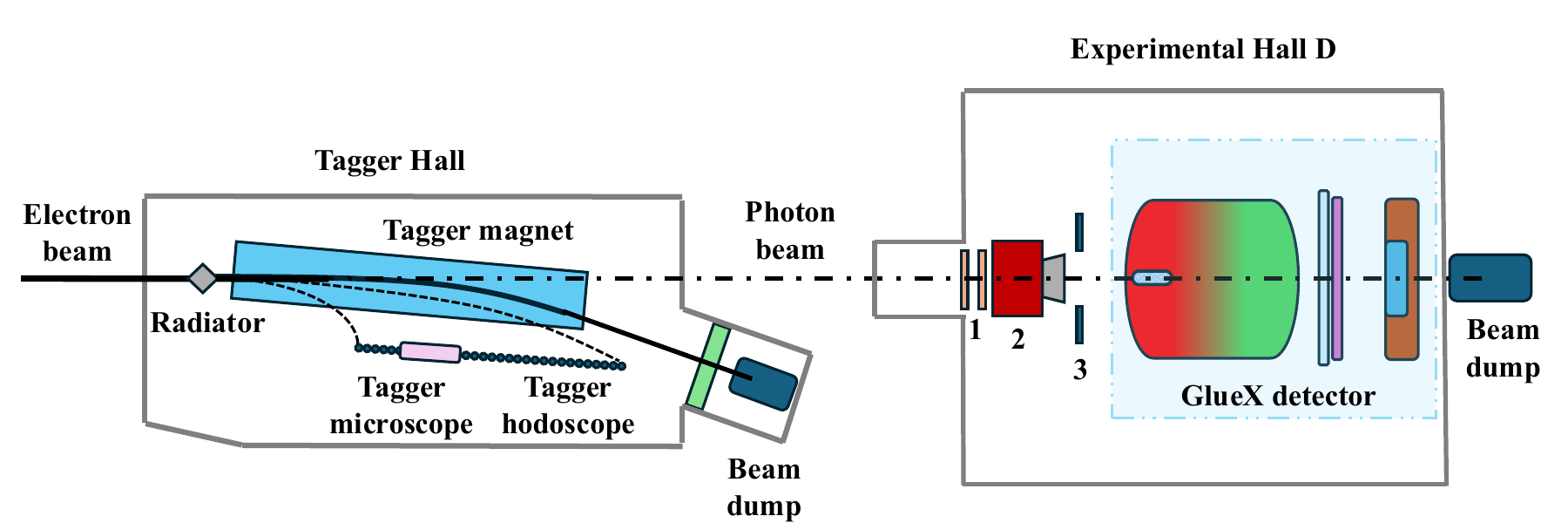}
\end{center}
\caption{Schematic view of the Hall~D photon beamline (not to scale). Numbers represent the following pair spectrometer components: converters (1), dipole magnet (2), detectors (3).}
\label{fig:gluex_beamline}
\end{figure*}

The photon beam in Hall~D is produced using a primary electron beam with a  typical energy of 12~GeV, delivered by the Continuous Electron Beam Accelerator Facility (CEBAF) at Jefferson Lab (JLab). Photons are generated via the bremsstrahlung process, in which the electrons interact  with a thin radiator inserted into the beamline.  When a diamond radiator is employed, linearly polarized photons are produced through coherent bremsstrahlung from the crystal lattice of the diamond. The polarized photons are emitted at specific energies, producing sharp, monochromatic  peaks superimposed on the broader bremsstrahlung continuum of the photon energy spectrum. The energies of these peaks, as well as the degree of linear polarization, depend on the orientation of the crystal lattice relative to the incident electron beam. The fraction of linearly polarized photons is enhanced by passing the photon beam through a collimator that filters out non-polarized photons produced by incoherent bremsstrahlung at larger emission angles. In the nominal detector configuration, a collimator with a diameter of 5~mm is located approximately 75~m downstream of the radiator. The resulting collimated beam is incident on the target of the GlueX detector~\cite{gluex_det}. The GlueX detector is a forward magnetic spectrometer designed to carry out experiments using photon beams. It was commissioned in 2016 and is currently in operation, collecting experimental data. A schematic view of the Hall~D photon beamline is shown in Fig.~\ref{fig:gluex_beamline}.

The energy of a beam photon is determined by detecting the electron that emitted the photon via bremsstrahlung using a dedicated tagging system. After radiation, the scattered electron is momentum-dispersed by the magnetic field of a 6-meter-long dipole magnet and registered by two types of scintillator-based detectors with different segmentation: the broadband Tagger Hodoscope (TAGH) and the high-resolution Tagger Microscope (TAGM). Both detectors consist of arrays of scintillator counters, with each counter corresponding to a well-defined energy interval of the scattered (tagged) electron. The photon energy is obtained by subtracting the measured energy of the scattered electron from the known energy of the incident electron before radiation. This tagging system provides a typical relative photon energy resolution of approximately $0.1\%$.

A key component of the Hall D photon beamline is a magnetic Pair Spectrometer (PS), located approximately 10~m upstream of the GlueX target.  The PS reconstructs the energy spectrum and flux of the collimated photon beam by detecting  $e^\pm$ pairs produced when beam photons interact in a thin converter. The PS is integrated into the GlueX trigger system through custom electronics developed at Jefferson Lab, enabling real-time monitoring of the photon flux and the position of the coherent peak in the photon energy spectrum. The flux information is essential for monitoring beam stability and for calibrating the active collimator~\cite{gluex_det}, a detector that provides feedback to the accelerator for steering the photon beam.

Triggers generated by $e^\pm$ pairs in the PS are recorded and used in the data analysis to determine the photon flux and, consequently, the luminosity required for measuring cross sections of various physics reactions. Electron-positron pairs reconstructed by the PS are also used to identify the triplet photoproduction events, in which an additional recoil electron is detected in the triplet polarimeter~\cite{polarimeter}. The polarization of the photon beam can be extracted from the azimuthal distribution of these recoil electrons.

%----------------------------------------------------------------------
%----------------------------------------------------------------------

Leptons with energies determined by the PS were used to test various calorimeter prototypes positioned downstream of the spectrometer. These tests were conducted concurrently with GlueX data taking. In particular, the PS test setup was extensively employed to optimize the design and study the performance of PbWO$_4$-based calorimeter modules intended for the upgrade of the GlueX forward calorimeter~\cite{ecal}. This article is organized as follows: the design of the pair spectrometer is described in Section~\ref{sec:ps}; its performance is presented in Section~\ref{sec:ps_perf}; and the calorimeter test setup based on the PS is discussed in Section~\ref{sec:ps_test_setup}.

%----------------------------------------------------------------------
\begin{figure}[t]
\begin{center}
\includegraphics[width=1.0\linewidth,angle=0]{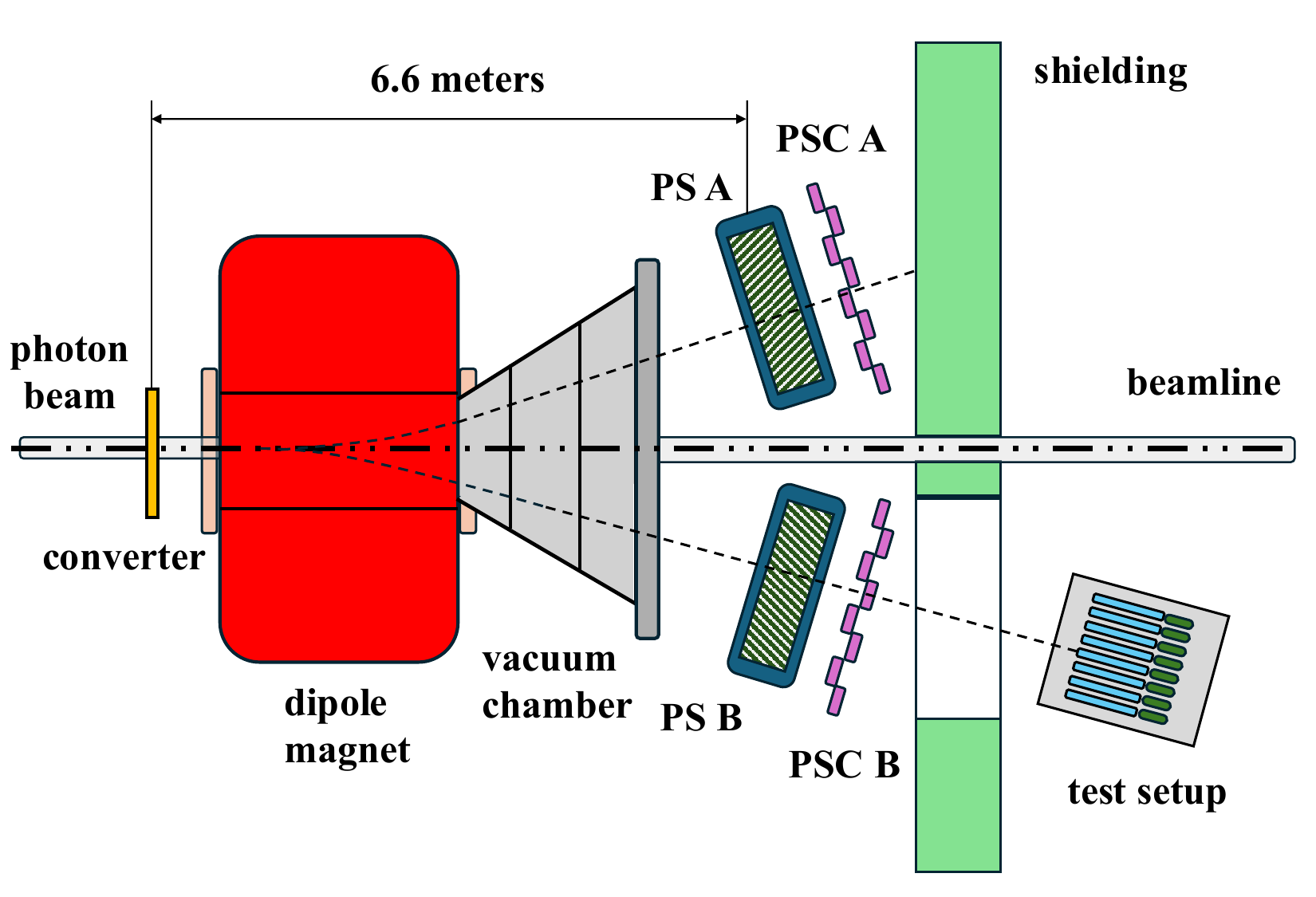}
\end{center}
\caption{Schematic view of the Hall~D pair spectrometer. The main components include the converter, dipole magnet, vacuum chamber, high-granularity hodoscopes (PS~A and PS~B), and coarse counters (PSC~A and PSC~B) in the two detector arms.}
\label{fig:gluex_ps}
\end{figure}
%----------------------------------------------------------------------

\section{Pair Spectrometer}
\label{sec:ps}

\begin{figure*}[t]
\begin{center}
\includegraphics[width=0.7\linewidth,angle=0]{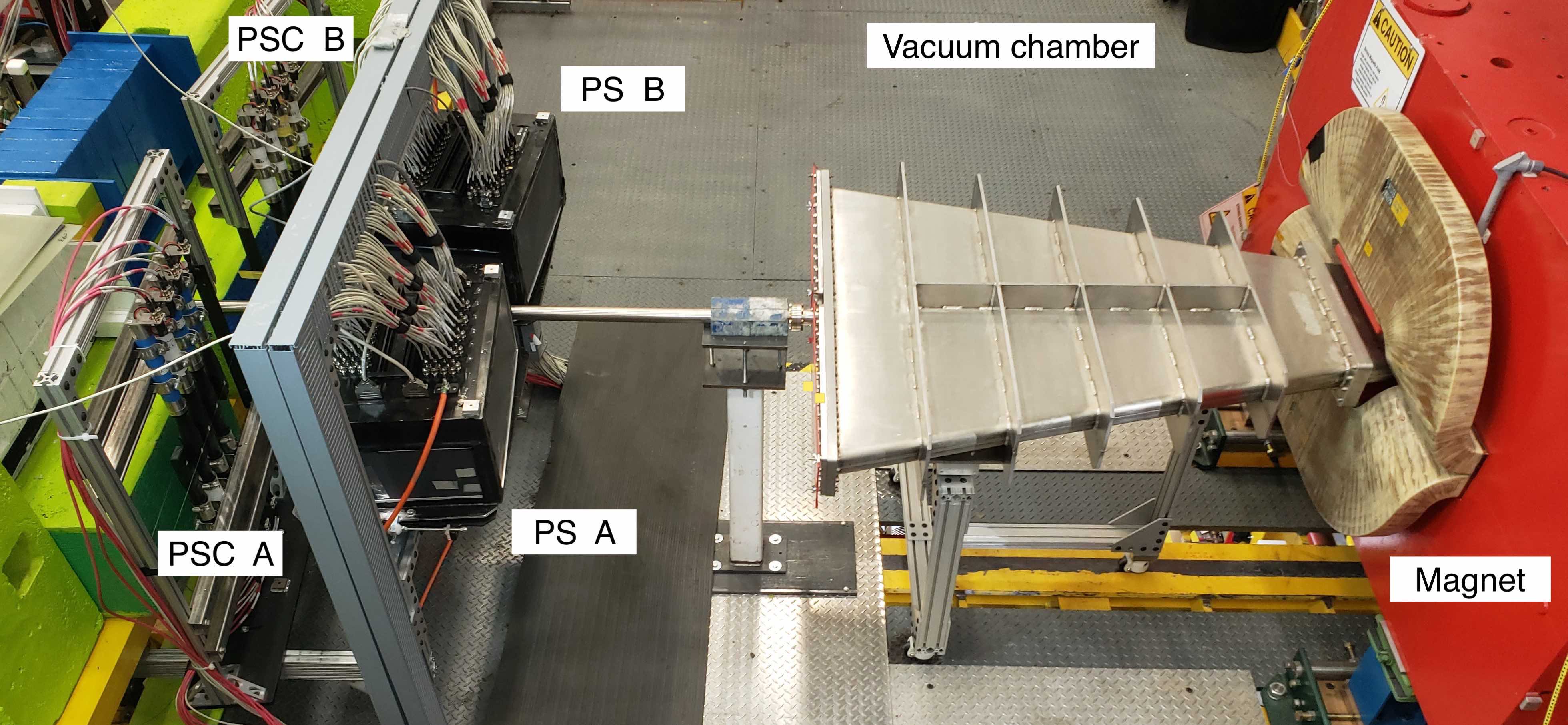}
\end{center}
\caption{Pair spectrometer installed in Hall~D.}
\label{fig:ps_photo1}
\end{figure*}

A schematic of the pair spectrometer is presented in Fig.~\ref{fig:gluex_ps}.  Electron-positron pairs are produced in a thin beryllium converter inserted into the photon beam. Two converters with thicknesses of $750\;{\rm \muup m}$ and $75\;{\rm \muup m}$ were used in the  Hall~D experiments. The converter thickness was selected based on the experimental conditions and the required photon flux. The converters are mounted on a remotely-controlled movable fixture, allowing one converter to be positioned in the beam. A dipole magnet with an effective field length of 94~cm, operating at magnetic fields up to 1.8~T, is used to deflect the leptons into the electron and positron arms of the spectrometer. To minimize interactions with air, a 1.5~m-long vacuum chamber is installed downstream of the magnet. Leptons are detected in each arm by two  scintillator-based detectors: the high-granularity hodoscope and the coarse counters. A photograph of the pair spectrometer installed in Hall~D is shown in Fig.~\ref{fig:ps_photo1}.

%----------------------------------------------------------------------
%----------------------------------------------------------------------
\begin{figure}[t]
\begin{center}
\includegraphics*[width=0.9\linewidth,angle=0]{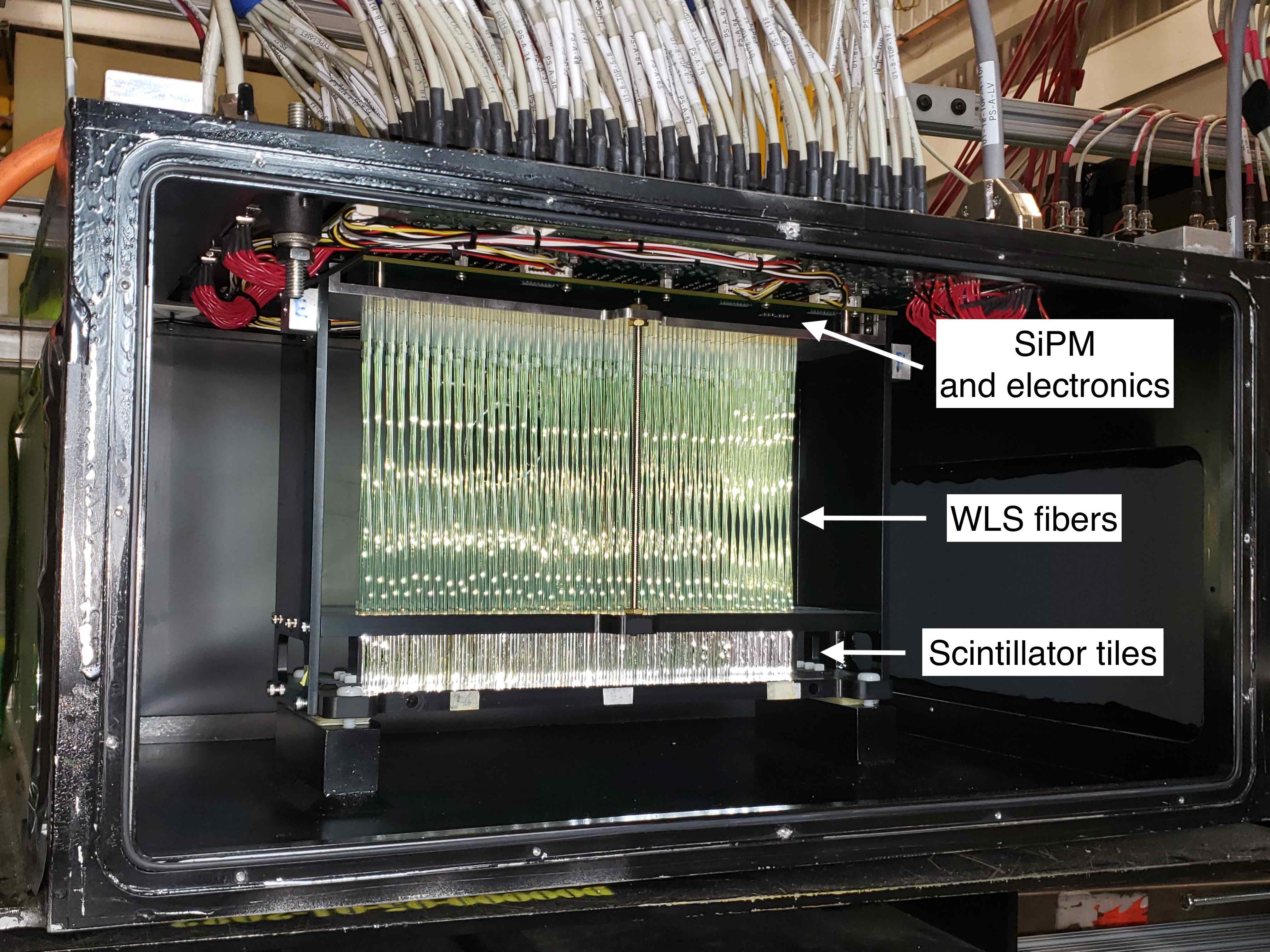}
\end{center}
\caption{Pair spectrometer hodoscope with associated electronics installed inside the light-tight enclosure.}
\label{fig:ps_photo2}
\end{figure}
%----------------------------------------------------------------------
\begin{figure}[t]
\begin{center}
\includegraphics*[width=0.7\linewidth,angle=0]{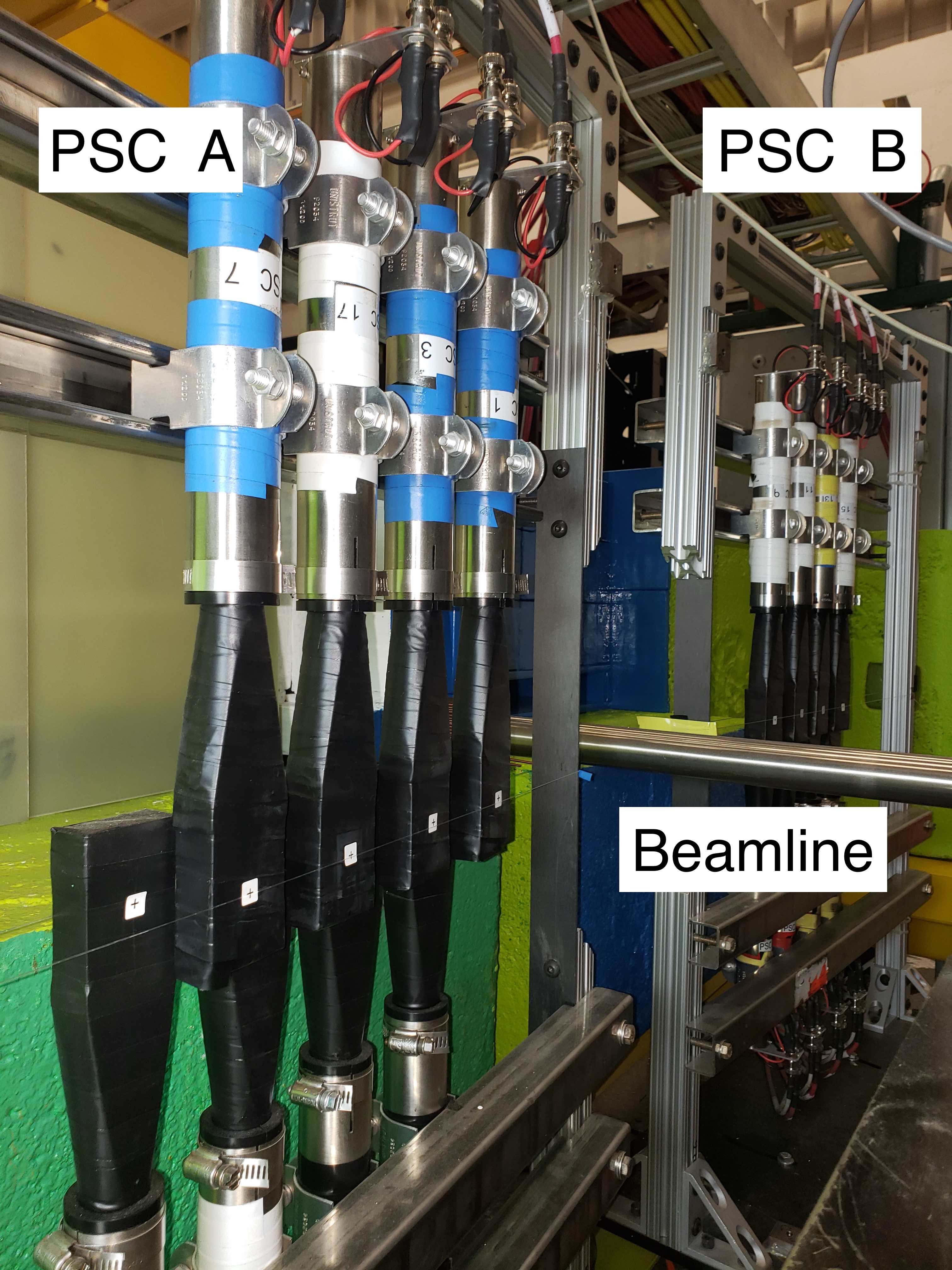}
\end{center}
\caption{Pair spectrometer coarse counters. Eight counters are installed in each spectrometer arm: the electron arm (PSC~A) and the positron arm (PSC~B).}
\label{fig:ps_photo3}
\end{figure}
%----------------------------------------------------------------------

\subsection{High-granularity hodoscope}
The high-granularity hodoscope is used to measure the momentum of leptons. Two identical detectors are installed symmetrically in the two arms of the spectrometer with respect to the photon beamline.  In the nominal configuration, the hodoscope covers a lepton energy range of approximately 3-6~GeV. Each detector consists of 145 thin rectangular scintillator tiles made of EJ-212 (Eljen Technology) stacked along the lepton trajectory. The tiles are 1~cm long and 3~cm high, while their width varies with energy. In the low-energy region, tiles are 2~mm wide, whereas in the high-energy region near 6~GeV, 40 tiles with a reduced width of 1~mm are installed to improve energy resolution. Scintillation light from each tile is collected by two square wavelength-shifting (WLS) fibers (BCF-92), each with a cross section of $1\;{\rm mm} \times 1\;{\rm mm}$. The fibers are glued to opposite sides of the tile using BC~600 optical cement. Light is transported over 20~cm and detected by a single Hamamatsu S10931-050P silicon photomultiplier (SiPM) coupled to both fiber ends. The SiPMs have a sensitive area of $3\;{\rm mm}\:{\times}3\;{\rm mm}$ and a pixel pitch of $50\;{\rm \muup m}$.  The front-end electronics include a SiPM bias voltage control system and an amplifier with a gain of approximately 20. The SiPMs operate at a bias voltage of approximately 73~V supplied by ISEG EHS voltage modules. Low-voltage power ($\pm 5\;{\rm V}$) for the front-end electronics is provided by an MPV8008 MPOD module. Both power supply modules are housed in a Wiener crate. Signal pulses are digitized using 12-bit, 16-channel flash Analog-to-Digital Converter (ADC) modules~\cite{fadc250}  operating  at a sampling rate of 250~MHz. These ADCs, designed at Jefferson Lab, are also used for several GlueX sub-detectors. A design of the PS hodoscope is described in Ref.~\cite{ps_hod}. Figure~\ref{fig:ps_photo2} shows one hodoscope arm together with the associated electronics installed inside the light-tight enclosure.

%----------------------------------------------------------------------
\begin{figure}[htb]
\begin{center}
\includegraphics*[width=1.0\linewidth,angle=0]{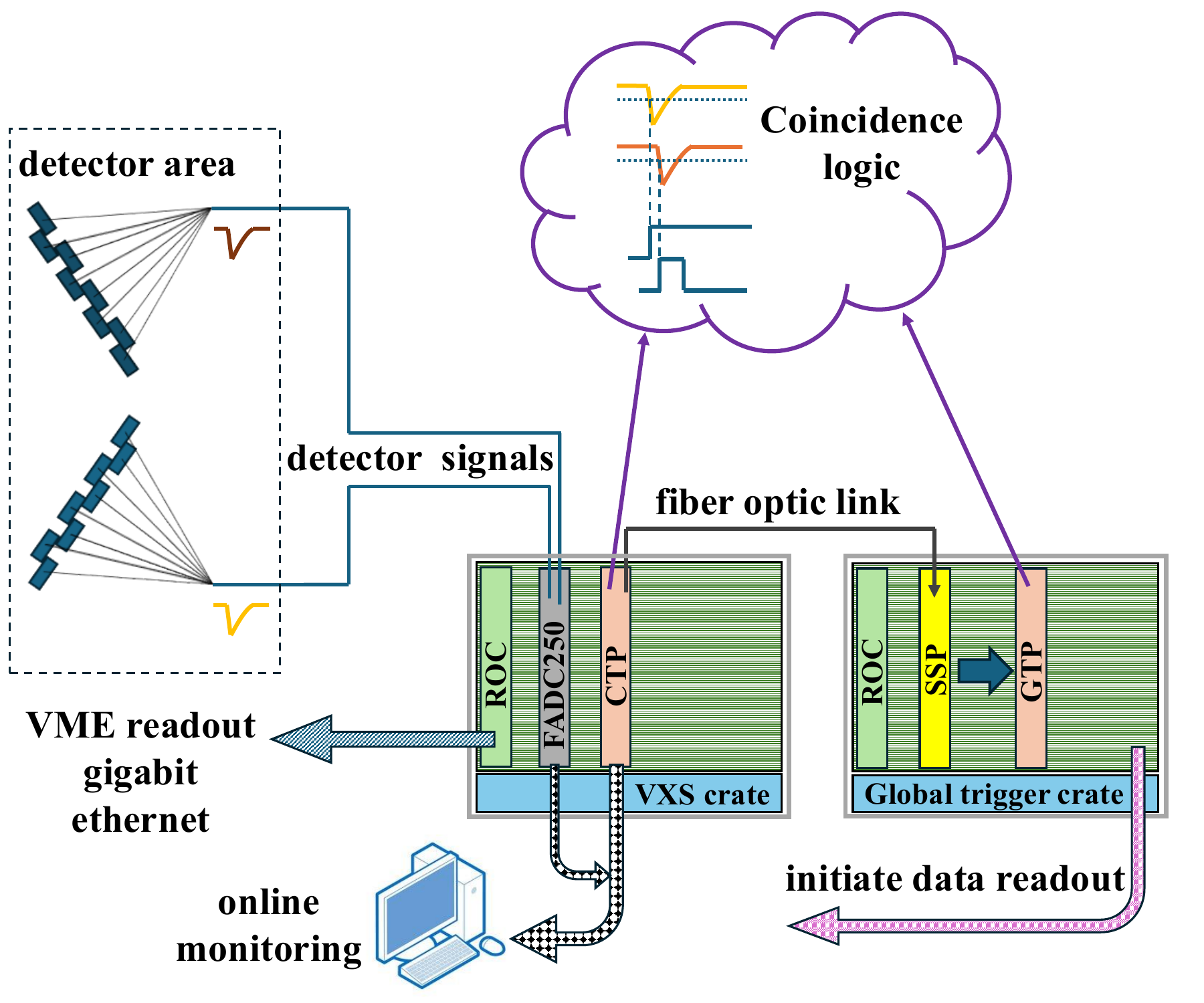}
\end{center}
\caption{Schematic view of the pair spectrometer trigger and monitoring system.}
\label{fig:ps_trigger}
\end{figure}
%----------------------------------------------------------------------

%------------------------------------------------
\begin{table*}[htb]
\begin{center}
\renewcommand{\arraystretch}{1.4}
\begin{tabular}{l|c|c|c|c|c}
\hline
Conditions &  \multicolumn{4}{c}{Experiments}   \\  \cline{2-6}
                         & GlueX~\cite{gluex1} & PrimEx~\cite{primex} & SRC~\cite{src} & CPP/NPP~\cite{cpp,npp} & GlueX II/JEF~\cite{,gluex2,jef}\\ 
\hline
\hline
$\rm E_{Beam}$ (GeV)  & 11.6  & 10, 11.2, 11.6 & 10.8 & 11.6  &  11.7 \\
\hline
PS field (T)  & 1.64 & 1.56 & 1.56 & 1.06 & 1.66 \\
\hline 
PS energy range (GeV) & 6.2 - $E_{\rm beam}$  & 5.9 - $E_{\rm beam}$ & 5.9 - $E_{\rm beam}$ & 4.1 - 7.8 & 6.3 - $E_{\rm beam}$  \\
\hline
PS converter thickness ($\rm \muup m$) & 75 & 750 & 75 & 750 & 75 \\
\hline
Photon flux ($\gamma$/sec) & $2\cdot10^{7}$  & $0.9\cdot10^{7}$ & $2\cdot10^{7}$ & $1.1\cdot10^{7}$ &  $5\cdot10^{7}$ \\
$\rm E_{Range}$ (GeV)  & 8.4 - 9.0  & 8 - $E_{\rm beam}$ & 8.0 - 9.0 & 5.5 - 6.0 & 8.4 - 9.0 \\
\hline
Beam polarization & yes & no & yes & yes & yes\\
\hline
\end{tabular}
\end{center}
\caption{Summary of the run conditions and pair spectrometer settings for the Hall~D experiments. $E_{\rm Beam}$ denotes the electron beam energy, and the photon flux corresponds to the typical flux in the energy range of interest, $E_{\rm Range}$.} 
\label{tab:halld_exp}
\end{table*}
%------------------------------------------------

\subsection{Coarse counters}

Eight scintillator counters are installed in each PS arm, approximately 40~cm downstream of the hodoscope along the beam line. These counters are used in the GlueX trigger system and provide timing information to identify the accelerator beam bunch associated with an interaction in the PS. This timing allows correlation of hits in the tagging detectors and the pair spectrometer originating from the same event. Each counter consists of a rectangular EJ-200 plastic scintillator with  dimensions of 2~cm along the lepton trajectory, 4.4~cm perpendicular to the  trajectory, and 6~cm in height. Scintillation light is detected by a Hamamatsu R6427-01 photomultiplier tube (PMT) coupled to the scintillator via a fishtail light guide. The light guide is optically glued to both the scintillator and the PMT entrance window using a Dymax UV-cured optical adhesive. The PMTs are typically operated at a high voltage of 1250~V, supplied by a CAEN A1535SN high-voltage module installed in a SY1527 mainframe. The PMT signal is split into two paths: one is sent to a flash ADC for waveform digitization and the trigger processing, while the other is fed into a leading-edge discriminator followed by a Time-To-Digital Converter (TDC) for precise timing measurements. The TDC, designed at Jefferson Lab, has a bin width of approximately 58~ps. The installed coarse counters are shown in Fig.~\ref{fig:ps_photo3}.

\subsection{Pair spectrometer trigger and online monitoring}

The pair spectrometer trigger is generated by leptons originating from $e^+e^-$ pairs. It requires time-correlated hits in the coarse PS counters located in both the electron and positron arms of the spectrometer. The trigger logic is implemented using special-purpose programmable electronics modules equipped with Field-Programmable Gate Array (FPGA) chips. A detailed description of the GlueX trigger architecture is provided in Ref.~\cite{l1_trigger}.

Trigger processing begins at the flash ADCs. Sixteen coarse PS counters are connected to a single flash ADC module installed in a VXS crate compliant with the ANSI/VITA 41.0 standard. Signal processing is divided into two parallel paths: a trigger path and a data acquisition (DAQ) path. In the DAQ path, digitized waveforms are analyzed to extract pulse parameters including peak amplitude, integral, and timing. These quantities are stored in the pipeline memory for subsequent readout.

In the trigger path, the flash ADC identifies pulses whose amplitudes exceed a predefined threshold, typically 24~mV, and converts them into digital hit bits. This hit-bit information is transmitted via the VXS high-speed serial bus to the Crate Trigger Processor (CTP) located at the center of the VXS crate. The CTP forwards the hit pattern via an optical link to the Sub-System Processor (SSP) in the trigger crate. Although the SSP can accept and process inputs from multiple crates, for the PS trigger it simply relays the hit pattern to the Global Trigger Processor (GTP). The GTP performs a coincidence search between hits in the two detector arms within a programmable time window, typically 20~ns wide, and generates the final trigger decision. Events satisfying the PS trigger are read out from the flash ADCs by the Readout Controller (ROC) and recorded for offline analysis.

In addition to event triggering, the PS trigger electronics provide continuous monitoring of the $e^+e^-$ coincidence rate as well as the individual detector hit rates, independent of the GlueX data-acquisition system. This functionality enables real-time monitoring of the photon flux. As in the GTP processing, the hit coincidence search between the electron and positron arms is performed at the crate level inside the CTP. The rates are computed in the FPGA firmware of both the CTP and flash ADC modules and are read out asynchronously from the detector DAQ via the ROC by a dedicated monitoring process. The monitoring data are distributed to online monitoring programs and archived in a database using the Experimental Physics and Industrial Control System (EPICS)~\cite{epics}. A schematic overview of the PS trigger system is shown in Fig.~\ref{fig:ps_trigger}.

%----------------------------------------------------------------------
\begin{figure*}[t]
\begin{center}
\includegraphics[width=1.0\linewidth,angle=0]{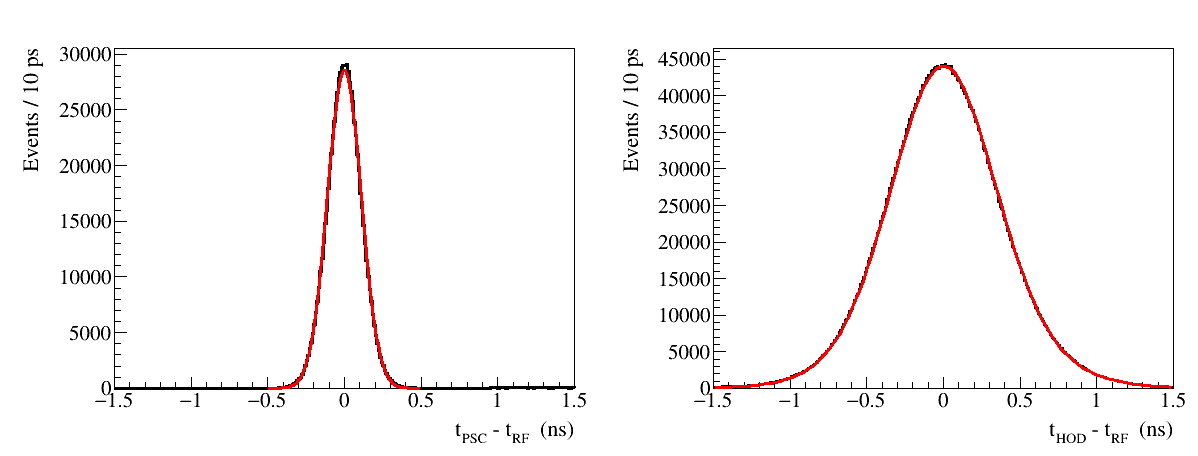}
\end{center}
\caption{Time difference between the pair spectrometer coarse counter (PSC) and the accelerator RF (left), and between the PS hodoscope (HOD) and the RF (right). The solid lines show fits using Gaussian functions, as described in the text.}
\label{fig:psc_ps_time}
\end{figure*}
%----------------------------------------------------------------------

\section{Performance}
\label{sec:ps_perf}

The pair spectrometer has been successfully used in multiple Hall~D experiments, beginning with GlueX, whose primary physics goal was the study of light-quark meson spectroscopy and the search for gluonic excitations using linearly polarized photons \cite{gluex1}. Each Hall~D experiment required specific photon beam conditions, including the beam polarization, the position of the bremsstrahlung coherent peak, the photon energy range, and the photon flux. To optimize the PS energy coverage, the magnetic field of the PS dipole was adjusted according to the requirements of each experiment. The corresponding run conditions and PS settings for various experiments are summarized in Table \ref{tab:halld_exp}.

During the recent GlueX~II/JEF experiments~\cite{gluex2,jef}, the detector operated at the highest photon flux. The PS trigger rate remained below 7.5~kHz, corresponding to approximately $10\%$ of the total experiment trigger rate. The maximum rates observed in the PS hodoscope and coarse counters were approximately 400~Hz and 4.5~kHz, respectively.

\subsection{Time resolution and lepton detection efficiency}

Timing measurements in the GlueX detector are performed relative to the accelerator radio-frequency (RF) signal of $\sim499$~MHz. A prescaled, low-jitter RF signal is distributed to several TDC channels of the GlueX sub-detectors as a reference time, enabling resolution of the beam-bunch structure with an effective period of approximately 4~ns. PS timing is measured using the TDC module of the coarse counters. To correct the amplitude-dependent time shift introduced by the leading-edge discriminator, a time-walk correction is applied using pulse amplitudes recorded by the flash ADC. The time difference between the PS counter and RF reference signals is shown in the left plot of Fig.~\ref{fig:psc_ps_time}, with a Gaussian fit superimposed. The typical time resolution of all PS counters is approximately 110~ps, which is sufficient to clearly separate events originating from different beam bunches. Accurate beam-bunch identification is essential for matching PS hits with those in the tagging detectors.

The PS hodoscope is not intended to provide precise timing information. The counters are not equipped with dedicated TDC modules; instead,  timing information is obtained from a processing algorithm implemented in the flash ADC FPGA. The algorithm identifies the pulse peak, the beginning of the pulse, and performs a linear interpolation between two consecutive samples around a fixed fraction of the pulse leading edge. The typical time resolution measured in one detector arm is approximately 350~ps, as shown in the right plot of Fig.~\ref{fig:psc_ps_time}. To account for the non-Gaussian tails in the timing distribution, the data were fitted with the sum of two Gaussian functions. For the events with hits in adjacent scintillator tiles, the final time is computed as an amplitude-weighted average of the individual tile times.

The detection efficiency of leptons in the PS  hodoscope was evaluated by using electrons and positrons from reconstructed $e^\pm$ pairs. One lepton was required to be fully reconstructed, with coincident hits in both the PS coarse counter and the corresponding hodoscope counters. The other lepton was required to have a hit in the coarse counter, and the efficiency was determined by searching for an associated hit in the hodoscope. The average efficiency was found to exceed $95\%$, with some inefficiency attributable to small gaps between scintillator tiles. The average light yield from a tile corresponds to approximately 60–90 SiPM pixels. For leptons detected with the coarse counters, the detection efficiency is nearly $100\%$.

\subsection{Energy resolution}

%----------------------------------------------------------------------
\begin{figure}[ht]
\begin{center}
\includegraphics[width=1.0\linewidth,angle=0]{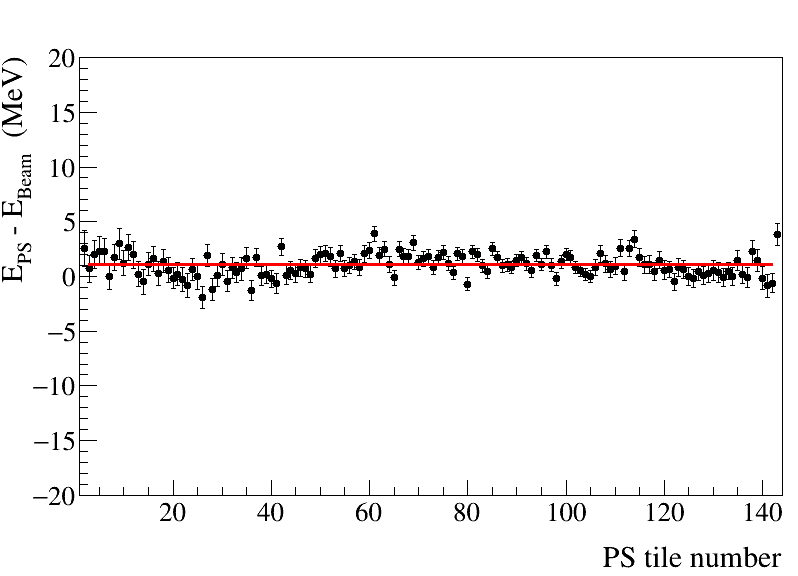}   
\end{center}
\caption{Energy residual, $\Delta E = E_{\rm PS} - E_{\rm Beam}$, between the PS-measured energy and the beam energy, shown as a function of the PS hodoscope tile number in the electron arm. The beam energy is $E_{\rm Beam} = 8.9$~GeV. The solid line represents a fit to the distribution using a zero-order polynomial (constant) function.}
\label{fig:ps_calib}
\end{figure}
%----------------------------------------------------------------------
\begin{figure}[htb]
\begin{center}
\includegraphics[width=1.0\linewidth,angle=0]{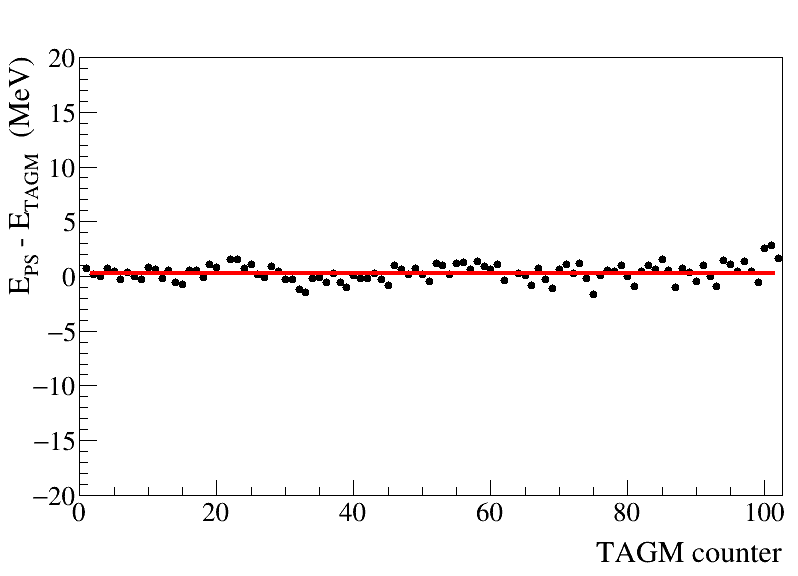}
\end{center}
\caption{Difference between the beam-photon energy measured by the pair spectrometer and the tagger microscope, plotted as a function of the TAGM counter number. The solid line shows a fit to a zero-order polynomial. }
\label{fig:ps_tagm_dep}
\end{figure}
%----------------------------------------------------------------------
\begin{figure}[htb]
\begin{center}
\includegraphics[width=1.0\linewidth,angle=0]{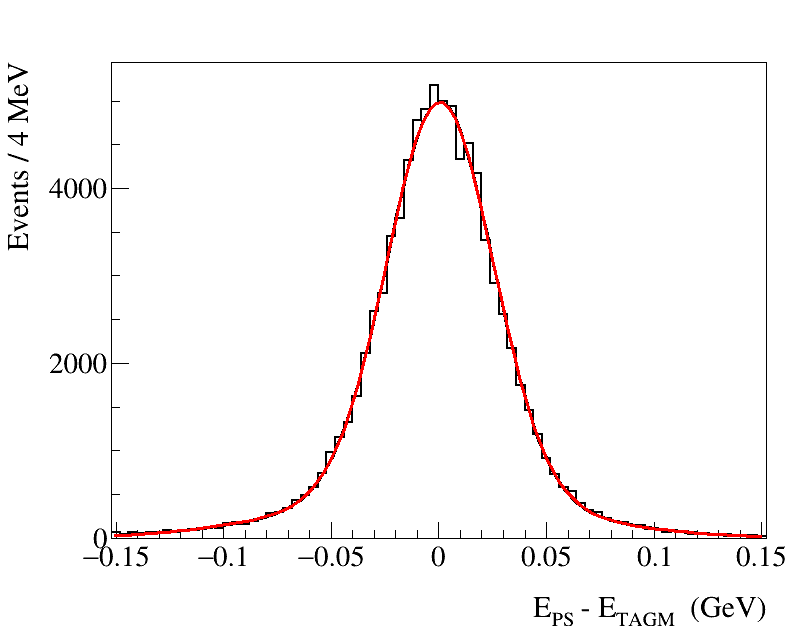}
\end{center}
\caption{Difference between the beam-photon energy reconstructed by the tagger microscope counter and the pair spectrometer. The solid line represents a fit to the distribution using the sum of two Gaussian functions. }
\label{fig:ps_tagm}
\end{figure}
%----------------------------------------------------------------------

The energy range of each counter in the pair spectrometer hodoscope was initially determined using a Geant simulation~\cite{geant4} incorporating a realistic geometry of the GlueX detector. Leptons originating from $e^+e^-$ pairs produced inside the PS converter were propagated through the pair spectrometer magnetic field to the PS counters. A three-dimensional magnetic field map of the dipole magnet was generated using the TOSCA program~\cite{tosca}, which includes the detailed 3D geometry of the magnet and the magnetic properties of its materials. This field map was then implemented in the Geant simulation. Magnetic field maps were produced for several field strengths.

To validate the TOSCA-generated magnetic field map, the magnetic field, $B$, was measured with a Hall probe in several regions of the magnet. The measurements were found to be in good agreement with the simulation: the integrated field along the lepton trajectory near the beamline, $\int {\rm B}\cdot d{\rm l}$, where $l$ is the path length, differed by only $4\cdot 10^{-4}$. This difference corresponds to a change in the reconstructed electron energy smaller than 3~MeV. The magnetic field of the PS magnet is continuously monitored using a Nuclear Magnetic Resonance (NMR)  field probe. The field stability was maintained at the level of $<10^{-4}$.

The energy scale of the PS hodoscope counters can be verified and, if necessary, adjusted using experimental data. Photons with known beam energy, defined by the tagging detectors, are selected and their energy is reconstructed using the pair spectrometer. The photon energy, $E_{\rm Beam}$, can be obtained from different combinations of counters in the electron and positron arms of the PS hodoscope. For a correctly calibrated detector, the energy residual, $\Delta E = E_{\rm e^-} + E_{\rm e^+} - E_{\rm Beam}$, is independent of the individual counter energies, $E_{e^\pm}$, i.e., it is flat as a function of $E_{e^\pm}$. This behavior was confirmed with Monte Carlo simulations. An example of the energy residual distribution for an unpolarized photon beam with $E_{\rm Beam}=8.9$~GeV is shown in Fig.~\ref{fig:ps_calib}. Any uncertainty in the beam energy does not change the shape of the residual distribution, affecting only the overall offset of $\Delta E$. We note that polarized photons affect the angular distribution of leptons produced in $e^+e^-$ pairs, causing them to be emitted along specific planes relative to the polarization direction. Small azimuthal effects in pair production by polarized photons can introduce a slight bias in the PS reconstructed energy, on the order of the PS energy resolution. These effects are negligible for the flux determination discussed in Section~\ref{sec:ps_flux}.

A comparison of the beam energy measured by the PS and the TAGM is shown in Fig.~\ref{fig:ps_tagm_dep}. The mean energy difference, $E_{\rm PS} - E_{\rm TAGM}$, is plotted as a function of the TAGM counter number. No significant bias is observed between the energies reconstructed by the two detectors. An example of the energy residual distribution for a single TAGM counter is shown in Fig.~\ref{fig:ps_tagm}, together with a fit using the sum of two Gaussian functions. The energy resolution, defined as the width of the core Gaussian component, is approximately 26~MeV. This resolution is dominated by the PS, since the intrinsic resolution of the TAGM counters is about 7~MeV. For the nominal magnetic field settings, the PS energy resolution varies from around 20~MeV to 29~MeV for beam-photon energies between 6~GeV and 12~GeV.

\subsection{Photon flux determination}
\label{sec:ps_flux}

One of the primary tasks of the PS is the determination of the photon flux incident on the GlueX target. Two types of photon flux are defined: untagged and tagged. The untagged flux corresponds to the total photon rate incident on the tagger and is primarily used for beam monitoring and for establishing stable running conditions. It is determined by counting the number of photons reconstructed in the PS within each beam-energy bin. Photons incident on the target may or may not be associated with a corresponding hit in the tagger detectors, so their energies are not always known. An example of the untagged flux spectrum is presented in Section~\ref{sec:beam_spectrum}.

In contrast, the tagged photon flux comprises photons whose energies are measured by the tagger detectors. This flux is required for absolute cross section determinations and relies on the precise photon-energy calibration provided by the TAGH and TAGM detector systems. The tagged flux extraction is performed independently for each photon-energy bin corresponding to individual TAGH and TAGM counters  and therefore requires a reconstructed beam photon in coincidence with the corresponding PS $e^+e^-$ pair.

\begin{figure}[htb]
\begin{center}
\includegraphics[width=1.0\linewidth,angle=0]{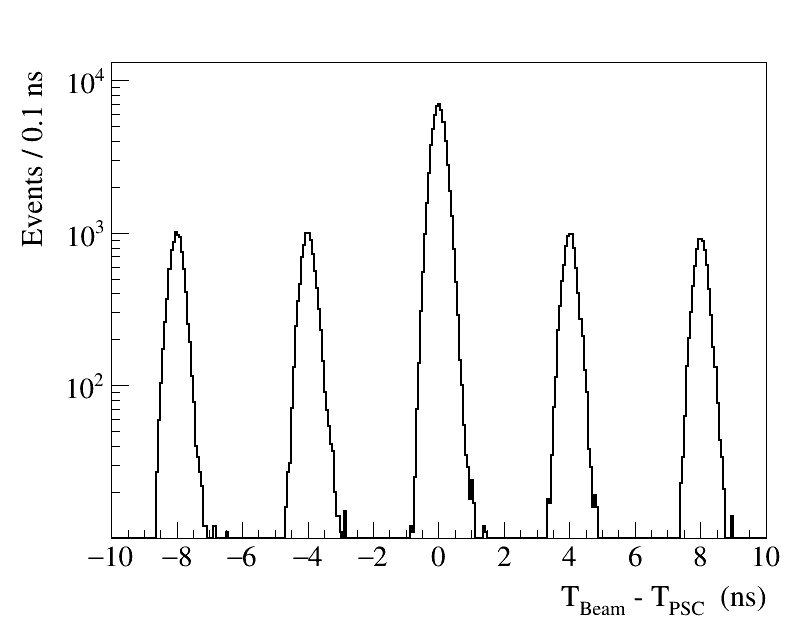}
\end{center}
\caption{Distribution of the time difference between beam photons detected by the TAGH counters and PS $e^+e^-$ pairs measured by the coarse counters. The peaks are separated by the beam bunch period of 4~ns.}
\label{fig:beam_bunch}
\end{figure}

Under typical Hall~D running conditions, the tagger detectors exhibit relatively high  hit multiplicities, resulting in a substantial rate of accidental time coincidences between reconstructed $e^+e^-$ pairs in the PS and hits in the tagger detectors. This background is suppressed by requiring an energy coincidence: only tagger hits with energies within a typical window of $\pm200$~MeV relative to the photon energy reconstructed in the PS are accepted. The distribution of the time difference between beam-photons candidates reconstructed by the TAGH and the corresponding PS pairs, after requiring the energy-coincidence, is shown in Figure~\ref{fig:beam_bunch}. The central peak corresponds to beam photons and $e^+e^-$ pairs originating from the same beam bunch, while the satellite peaks, separated by the $\sim$4~ns bunch period, represent accidental coincidences with photons from neighboring bunches that are not associated with interactions in the detector. The number of true coincidences is obtained by subtracting the accidental contribution, estimated from the satellite peaks, from the yield in the central peak.

 %----------------------------------------------------------------------
%----------------------------------------------------------------------
\begin{figure}[htb]
\begin{center}
\includegraphics[width=1.0\linewidth,angle=0]{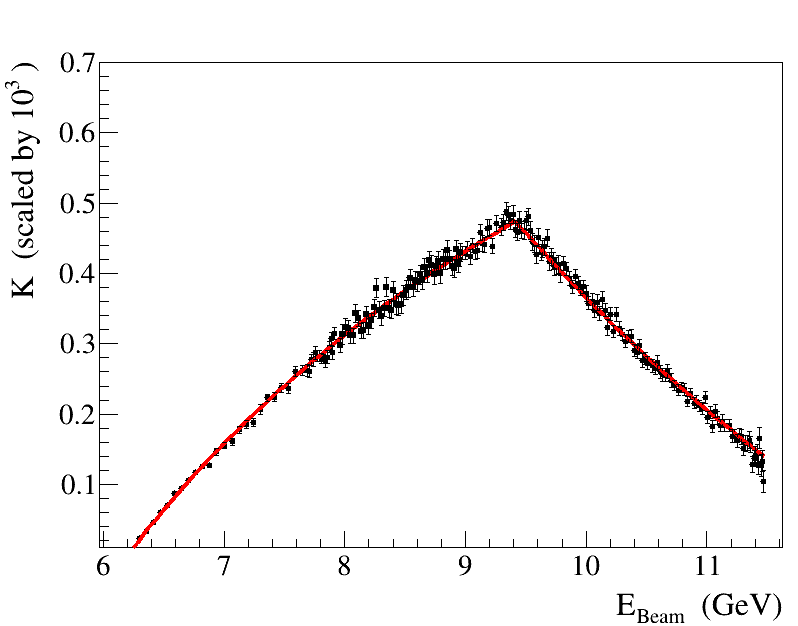}
\end{center}
\caption{Normalization coefficient $K$ obtained from the PS calibration run, shown as a function of the beam energy measured by the tagging detectors. The TAGM covers the energy range between 8-9~GeV. The solid line represents a fit using the function described in Eq.~(\ref{eq:fit_func}).}
\label{fig:ps_accept}
\end{figure}
%----------------------------------------------------------------------

The determination of the tagged photon flux using the pair spectrometer is described below. The number of beam photons, $N_{\gamma}$, associated with a given tagging counter is related to the number of reconstructed electron–positron pairs, $N_{e^+ e^-}$, by
%------------------------------------------------
\begin{equation}
N_{\gamma}  = \frac{N_{e^+e^-}}{\sigma_{e^+e^-} \cdot N_{\rm conv} \cdot L \cdot \epsilon \cdot A},
\label{eq:flux}
\end{equation}
%------------------------------------------------
where $\sigma_{e^+e^-}$ is the pair production cross section, $N_{\rm conv}$  is the number density of atoms in the PS converter, $L$ is the converter length, $\epsilon$ is the PS lepton detection efficiency, and $A$ is the PS acceptance.  The acceptance is defined as the probability that a produced $e^+e^-$ pair has leptons within the detection energy range of the PS.  The energy-dependent normalization factor in Eq.~\ref{eq:flux}, $K = \sigma_{e^+e^-}\: N_{\rm conv}\: L\:  \epsilon\; A$, was determined from dedicated PS calibration runs, in which the yields of reconstructed electron-positron  pairs and incident beam photons were measured simultaneously. For this purpose, a compact electromagnetic calorimeter, referred to as the total absorption counter (TAC), was inserted into the photon beam to provide a direct measurement of the photon rate. The calorimeter was mounted on a movable platform located approximately 12~m downstream of the GlueX target. In early Hall~D experiments, the TAC consisted of a lead-glass block; it was subsequently replaced by a calorimeter composed of lead tungstate modules, providing improved radiation hardness. 

During the PS calibration runs, data were collected using two triggers operating in parallel: (1) the standard PS trigger, which selected $e^+e^-$ candidates, and (2) the TAC trigger, based on the energy deposited in the calorimeter, which accepted beam-photon candidates. For a given photon energy, the numbers of detected $e^+e^-$ pairs, $N_{e^+e^-}^{\prime}$, and TAC photon candidates, $N_{\rm TAC}$, are related to the number of incident tagged photons, $N_\gamma$, as 
\begin{align}
N_{e^+e^-}^{\prime} & =  N_{\gamma} \cdot K  \notag \\
N_{\rm TAC} & =  \frac{N_{\gamma} \cdot \delta}{P},
\label{eq:calib}
\end{align}
where $P$ is the TAC trigger prescale factor, introduced to maintain the data acquisition livetime close to $100\%$, and  $\delta$ is the TAC photon detection efficiency. Small inefficiencies  due to interactions of beam photons with beamline material were evaluated using a detailed Geant simulation and  found to be $0.3\%$. The normalization factor $K$ can be derived from  Eq.~(\ref{eq:calib}) as
%------------------------------------------------
\begin{equation}
K  = \frac{N_{e^+e^-}^\prime \cdot \delta}{N_{\rm TAC}\cdot P}
\label{eq:calib1}
\end{equation}
%------------------------------------------------

The calibration runs were performed using low photon flux in order to minimize pileup effects in the TAC. Photons were produced by an electron beam with an equivalent current below 1~nA incident on the thinnest Hall~D aluminum radiator, with a thickness of $2\cdot 10^{-5}$ radiation lengths. The typical TAC hit rate, measured  at a 15~MeV energy threshold, was below 200~kHz. 

An example of the normalization factor $K$, measured using a $750\;{\rm \muup m}$ Be converter, is shown in Fig.~\ref{fig:ps_accept}  as a function of the beam-photon energy. The triangular shape arises from the PS acceptance, $A$, defined in Eq.~(\ref{eq:flux}), which is determined by the energy coverage of the PS electron and positrons arms. For this calibration run, the PS arms covered energies from approximately 3.2~GeV to 6.2~GeV. The acceptance depends on the number of possible $e^\pm$ energy combinations that can reconstruct a photon energy $E$ and can be parameterized as  
\begin{equation}
    A(E)=\begin{cases}
        N\cdot (1-\frac{2E_{\rm min}}{E}), & \text{if}\quad 2 E_{\rm min} < E \leq E_{\rm min}+E_{\rm max} \\
        N\cdot (\frac{2E_{\rm max}}{E}-1), & \text{if}\quad E > E_{\rm min}+E_{\rm max},
    \end{cases}
    \label{eq:fit_func}
\end{equation}
where $N$ is the normalization constant, and $E_{\rm min}$ and $E_{\rm max}$ are the minimum and maximum energies of the PS arms, respectively. The acceptance approaches zero at a photon-beam energy of approximately 6.4~GeV, corresponding to the lowest combination of the PS $e^\pm$ counters, whose individual energies are 3.2 GeV. It peaks near 9.4~GeV, where the number of reconstructible combinations is maximal. The upper limit of the acceptance, 11.4~GeV, corresponds to the maximum tagged-photon energy. The fit of the normalization factor using the parameterization in Eq.~(\ref{eq:fit_func}) is superimposed in Fig.\ref{fig:ps_accept}.

Due to the limited statistics of calibration runs, the normalization factor $K$ was parameterized; with sufficiently large statistics, the measured values for individual tagger counters could be used directly in Eq.~(\ref{eq:flux}). In the analyses, the fit function was further augmented with polynomial terms to account for the energy–angular dependence of the pair production cross section and the PS detection efficiency. The systematic uncertainties associated with the shape of $K$ typically dominate the overall systematic uncertainties in the flux determination. A recently published measurement of the Compton-scattering cross section by the PrimEx~$\eta$  experiment~\cite{primex} in the photon-energy range 6.5-11~GeV reports an overall systematic uncertainty of $3.4\%$~\cite{compton}. The uncertainties on $K$ range from approximately $1.5\%$ in the central region  of the PS energy acceptance to about $2\%$ near the acceptance edges, where the acceptance approaches zero. The systematic uncertainties associated with the flux shape can be reduced by acquiring a larger calibration data set. Throughout the experiment, the PS detector demonstrated stable performance, with an overall stability better than $1\%$.

\subsection{Photon beam spectrum}
\label{sec:beam_spectrum}

\begin{figure}[t]
\begin{center}
\includegraphics[width=1.0\linewidth,angle=0]{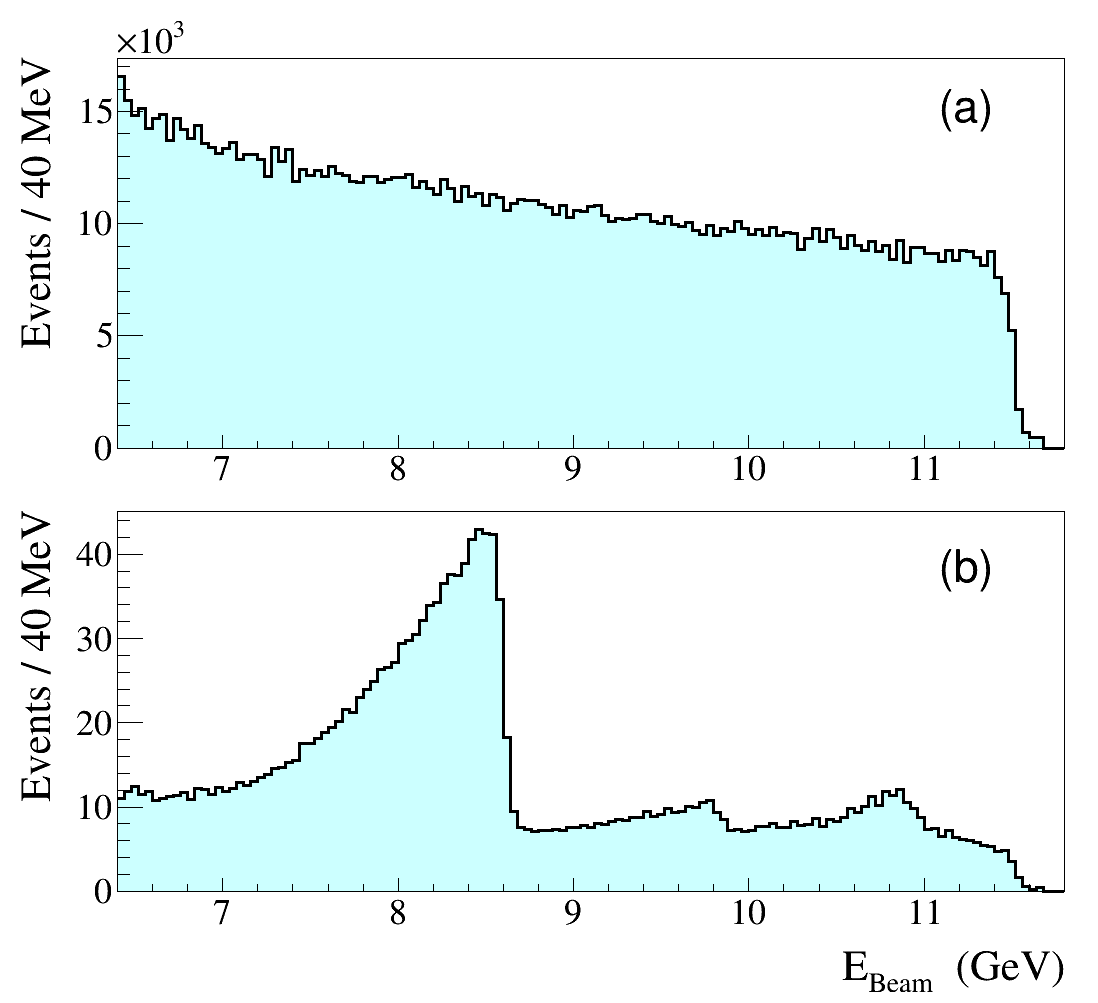}
\end{center}
\caption{Photon energy spectrum measured with the PS hodoscope for runs using (a) an amorphous radiator (unpolarized beam) and (b) a diamond radiator (linearly polarized beam with the main coherent peak around 8.6 GeV).}
\label{fig:ps_spect}
\end{figure}

The spectrum of the untagged beam photons incident on the GlueX target was obtained by reconstructing the photon energy in the pair spectrometer. Unlike the photon flux used for cross-section measurements, no coincidence with the tagging detector was required. The measured spectrum was corrected using the normalization factor defined in Eq.~(\ref{eq:flux}). The photon-beam spectrum obtained for runs with unpolarized photons produced by an amorphous (aluminum) radiator is shown in the upper plot of Fig.~\ref{fig:ps_spect}, while the lower plot shows an example measured with a diamond radiator. The peaks in the latter spectrum arise from linearly polarized photons produced via coherent bremsstrahlung in the diamond crystal lattice, as described in Section~\ref{sec:intro}. In the GlueX experiments, the position of the main coherent peak was monitored throughout data taking. In principle, the fraction of linearly polarized photons can be extracted from the the photon spectrum; this analysis is currently underway and will be reported in a forthcoming publication.

%----------------------------------------------------------------------
%----------------------------------------------------------------------

\section{Test setup using pair spectrometer}
\label{sec:ps_test_setup}

The pair spectrometer provided a unique capability for testing calorimeter prototypes using leptons produced via the $e^+e^-$ pair-production process. The test setup was installed in the positron arm of the PS, approximately 80~cm downstream of the coarse counters, as shown in Fig.~\ref{fig:gluex_ps}. This configuration allowed prototype tests to be conducted in parallel with GlueX data taking and enabled rapid reconfiguration during routine accelerator and detector maintenance.
 
Lepton energies were determined using the PS hodoscope counters, which spanned 3-6~GeV for most Hall~D experiments. For typical GlueX production running with a $75\;{\rm \muup m}$  Be converter and a 5~mm-diameter beam collimator, the relative energy resolution of leptons in the prototype region exhibited a weak energy dependence, increasing from approximately $0.5\%$ at 3~GeV to about $0.8\%$ at 6~GeV. The resolution depends on the beam-spot size on the PS converter, set by the collimator diameter. Since the $e^+e^-$ production vertex is not reconstructed in the PS, uncertainties in the vertex propagate directly into the reconstructed lepton energy. The relative energy resolution can be improved by using a smaller, 3.4~mm-diameter collimator. Although this collimator is available in the Hall~D beamline, it is not used during standard GlueX running. The vertical spread of  leptons in the test region is mostly determined by the primary collimator and is approximately 7~mm. Signals from most calorimeter prototypes were digitized using standard GlueX flash ADCs, integrated into the detector data acquisition and trigger systems. Offline analyses provided access to both PS and prototype data, allowing the reconstructed PS positron energy to be used in studies of prototype performance.

%----------------------------------------------------------------------
\begin{figure}[t]
\begin{center}
\includegraphics[width=1.0\linewidth,angle=0]{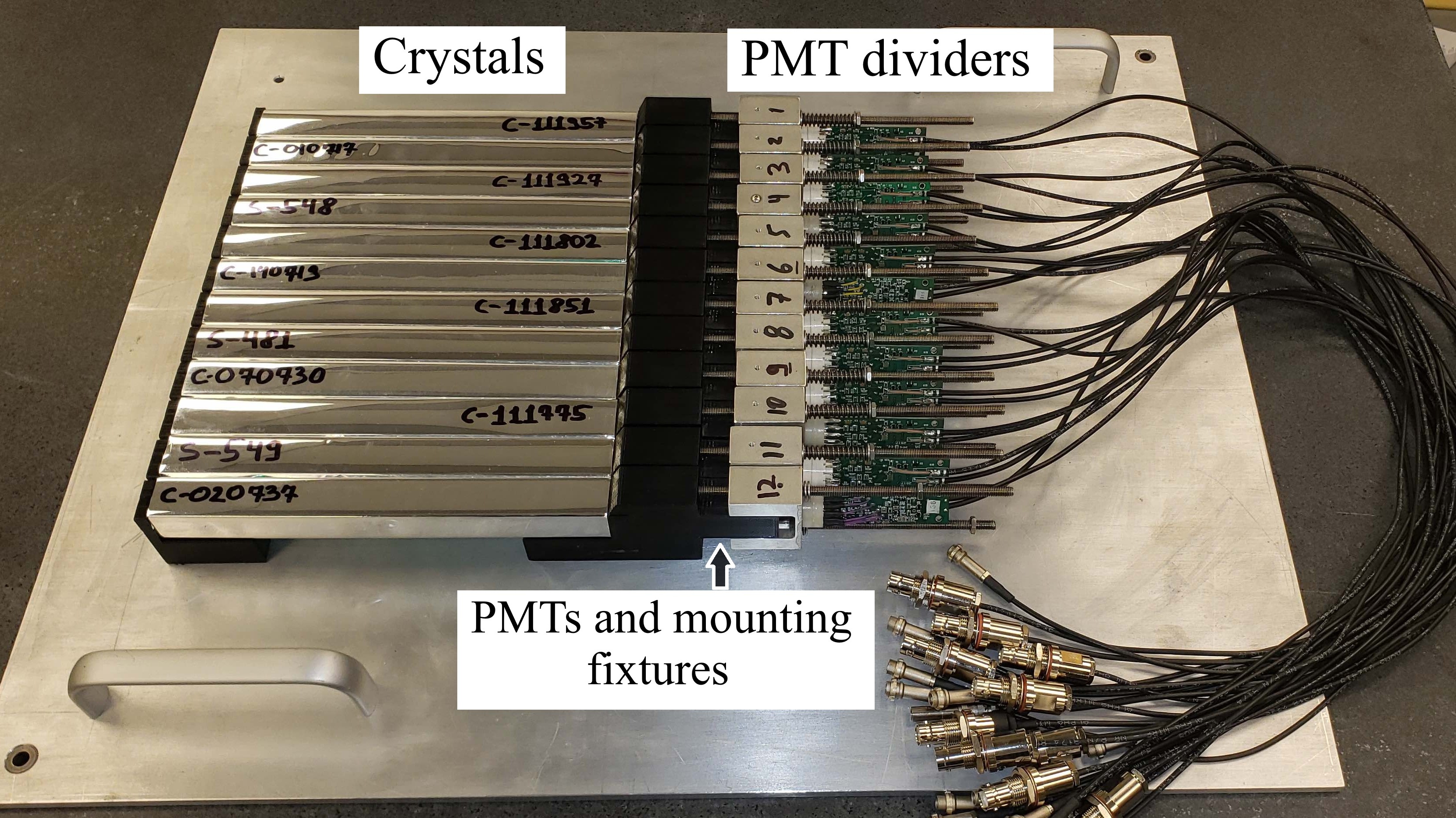}
\end{center}
\caption{Pair spectrometer setup used for testing PbWO$_4$ scintillating crystals for the ECAL calorimeter.}
\label{fig:ps_test_setup}
\end{figure}
%----------------------------------------------------------------------
 
\begin{figure}[t]
\begin{center}
\includegraphics[width=1.0\linewidth,angle=0]{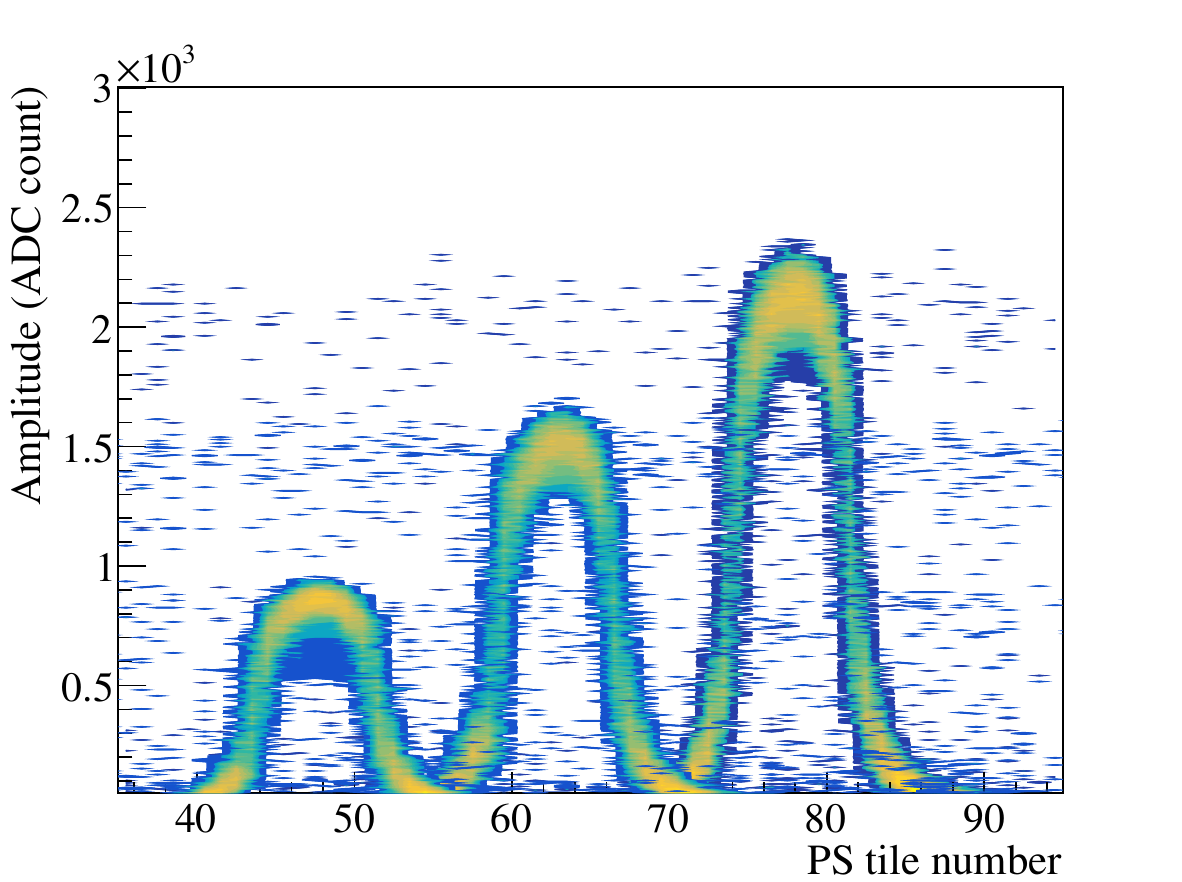}
\end{center}
\caption{Flash ADC amplitudes for three PbWO$_4$ calorimeter modules as a function of the pair spectrometer hodoscope tile number. Gaps indicate modules excluded from the analysis.}
\label{fig:ps_test_amp}
\end{figure}

The PS test setup was extensively used during the design and characterization of the GlueX lead tungstate electromagnetic calorimeter (ECAL)~\cite{ecal}, which consists of 1596 scintillating crystals and is required by the Jefferson Lab Eta Factory (JEF) experiment~\cite{jef}. The setup was initially employed to compare light yields of PbWO$_4$ crystals~\cite{pwo_crystals} sourced from two vendors: SICCAS (China) and CRYTUR (the Czech Republic), and was subsequently used for quality-assurance tests of newly acquired crystals.  Up to twelve crystals, each with dimensions  $2.05\;{\rm cm}\:{\times}\:2.05\;{\rm cm}\:{\times}\:20\;{\rm cm}$, were mounted on a plate, as shown in Fig.~\ref{fig:ps_test_setup}, and installed in the test setup to measure light yield and energy resolution. The plate was rotated by about $5^\circ$ relative to the beamline to ensure that positron trajectories were perpendicular to the crystal face at the center of the setup. Light from each crystal was detected using a Hamamatsu 4125 photomultiplier. An example of the flash ADC amplitudes for three PbWO$_4$ crystals as a function of the pair-spectrometer tile is shown in Fig.~\ref{fig:ps_test_amp}. The center of each plateau-shaped distribution corresponds to positrons passing through the middle of the crystal. The test setup was later adapted to study light collection for different crystal–PMT coupling configurations~\cite{ccal}.

Additional tests of PbWO$_4$ calorimeter prototypes included studies of silicon photomultipliers as alternative photodetectors~\cite{pwo_sipm}. More recently, a calorimeter prototype, consisting of an array of $3\times 3$ modules built from a new scintillating-glass material was investigated~\cite{scint_glass}. Further tests of calorimeter prototypes are ongoing, including evaluation of a hybrid lead-glass and PbWO$_4$ calorimeter. To improve the position resolution, a profiler detector is planned for installation in front of the prototypes.

%----------------------------------------------------------------------
%----------------------------------------------------------------------

\section{Summary}
\label{sec_summary}

The Pair Spectrometer in Hall D at Jefferson Lab demonstrated reliable performance across multiple experiments with the GlueX detector. The PS was used both to monitor the beam-photon flux on the target and, in combination with the tagging detectors, to determine the flux of tagged photons required for cross-section measurements. The achieved time resolution of the PS coarse counters, approximately 110~ps, enabled unambiguous identification of the accelerator beam bunch associated with each interaction. The energy resolution for reconstructed beam photons ranged from about 20~MeV at 6~GeV to 29~MeV at 12~GeV. In a recent measurement of the Compton-scattering cross section~\cite{compton}, the systematic uncertainty on the photon flux determination was approximately $2\%$. In addition, the PS served as a test facility, providing leptons with well-defined energies for the evaluation of various calorimeter prototypes in parallel with GlueX operation.

\section{Acknowledgments}
This material is based upon work supported by the U.S. Department of Energy, Office of Science, Office of Nuclear Physics under contract DE-AC05-06OR23177. The authors would like to thank Oleksandr Glamazdin for performing the simulation of the magnetic field of the PS magnet, and Alexandre Deur for conducting the field measurements.

\bibliographystyle{elsarticle-num}

\bibliography{ps_performance}

\end{document}